\documentclass[letterpaper]{article} % DO NOT CHANGE THIS
\usepackage{aaai25}  % DO NOT CHANGE THIS
\usepackage{times}  % DO NOT CHANGE THIS
\usepackage{helvet}  % DO NOT CHANGE THIS
\usepackage{courier}  % DO NOT CHANGE THIS
\usepackage[hyphens]{url}  % DO NOT CHANGE THIS
\usepackage{graphicx} % DO NOT CHANGE THIS
\usepackage{natbib}  % DO NOT CHANGE THIS AND DO NOT ADD ANY OPTIONS TO IT
\usepackage{caption} % DO NOT CHANGE THIS AND DO NOT ADD ANY OPTIONS TO IT
\usepackage{algorithm}
\usepackage{algorithmic}
\usepackage{newfloat}
\usepackage{listings}
\usepackage{array}
\usepackage{amsmath}
\usepackage{amssymb}
\usepackage{amsthm}
\usepackage{bm}
\usepackage{booktabs}
\usepackage{braket}
\usepackage{chngpage}
\usepackage{comment}
\usepackage{empheq}
\usepackage{enumitem}
\usepackage{longtable}
\usepackage{mathrsfs}
\usepackage{mathtools}
\usepackage[version=4]{mhchem}
\usepackage{ragged2e}
\usepackage{siunitx}
\usepackage{subcaption}
\usepackage{textcomp}
\usepackage{url}
\usepackage{verbatim}
\usepackage{vwcol}
\allowdisplaybreaks
\newcommand*\dif{\mathop{}\!\mathrm{d}}
\newcommand{\abs}[1]{\left|#1\right|}
\newcommand*\C{\mathop{}\!\text{core}}
\newcommand*\E{\mathop{}\!\text{edge}}
\newcommand*\SOL{\mathop{}\!\text{sol}}

\newcommand*\CE{\mathop{}\!\text{core-edge}}
\newcommand*\ES{\mathop{}\!\text{edge-sol}}
\newcommand*\SD{\mathop{}\!\text{sol-div}}
\newcommand*\R{\mathop{}\!\text{node}}
\newcommand*\ECR{\mathop{}\!\text{ECR}}
\newcommand*\NBI{\mathop{}\!\text{NBI}}
\newcommand*\SPI{\mathop{}\!\text{SPI}}
\newcommand*\ECH{\mathop{}\!\text{ECH}}
\newcommand*\ICH{\mathop{}\!\text{ICH}}
\newcommand*\LH{\mathop{}\!\text{LH}}
\newcommand*\GAS{\mathop{}\!\text{GAS}}
\newcommand*\RF{\mathop{}\!\text{RF}}
\newcommand*\IOL{\mathop{}\!\text{IOL}}
\DeclareCaptionStyle{ruled}{labelfont=normalfont,labelsep=colon,strut=off} % DO NOT CHANGE THIS
\floatstyle{ruled}
\newfloat{listing}{tb}{lst}{}
\floatname{listing}{Listing}
\usepackage{fancyhdr}
\fancypagestyle{firstpagehf}
{
    \fancyhf{}
    \fancyhead[HC]{The AAAI 2025 Workshop on AI to Accelerate Science and Engineering (AI2ASE)}
    \fancyfoot[C]{\thepage}
}
\nocopyright
\title{Application of Neural Ordinary Differential Equations for\\ITER Burning Plasma Dynamics}
\author {
    Zefang Liu\textsuperscript{\rm 1},
    Weston M. Stacey\textsuperscript{\rm 1}
}
\affiliations {
    \textsuperscript{\rm 1}Georgia Institute of Technology\\
    Atlanta, GA 30332, USA\\
    liuzefang@gatech.edu, weston.stacey@nre.gatech.edu
}
\begin{document}
%%%%%%%%%%%%%%%%%%%%%%%%%%%%%%%%%%%%%%%%
\thispagestyle{firstpagehf}
\maketitle
%%%%%%%%%%%%%%%%%%%%%%%%%%%%%%%%%%%%%%%%
\begin{abstract}
The dynamics of burning plasmas in tokamaks are crucial for advancing controlled thermonuclear fusion. This study applies the NeuralPlasmaODE, a multi-region multi-timescale transport model, to simulate the complex energy transfer processes in ITER deuterium-tritium (D-T) plasmas. Our model captures the interactions between energetic alpha particles, electrons, and ions, which are vital for understanding phenomena such as thermal runaway instability. We employ neural ordinary differential equations (Neural ODEs) for the numerical derivation of diffusivity parameters, enabling precise modeling of energy interactions between different plasma regions. By leveraging transfer learning, we utilize model parameters derived from DIII-D experimental data, enhancing the efficiency and accuracy of our simulations without training from scratch. Applying this model to ITER's inductive and non-inductive operational scenarios, our results demonstrate that radiation and transport processes effectively remove excess heat from the core plasma, preventing thermal runaway instability. This study underscores the potential of machine learning in advancing our understanding and control of burning plasma dynamics in fusion reactors.
\end{abstract}
%%%%%%%%%%%%%%%%%%%%%%%%%%%%%%%%%%%%%%%%
\section{Introduction}

The dynamics of burning plasmas in tokamaks are critical for advancing controlled thermonuclear fusion. In the International Thermonuclear Experimental Reactor (ITER) \cite{iaea2002iter}, deuterium-tritium (D-T) fusion reactions generate \SI{14.1}{MeV} neutrons and \SI{3.5}{MeV} fusion alpha particles \cite{green2003iter}. These alpha particles, confined by magnetic fields, transfer their energy primarily to core electrons before transferring it to ions. The heated core electrons emit various types of radiation, including electron cyclotron radiation (ECR), bremsstrahlung, and impurity radiation, which rapidly dissipate energy compared to the slower transport of energy to the edge. However, the energized electrons and remaining alpha particles also heat the core ions through collisional processes, potentially increasing fusion reactivity and leading to more fusion alpha particles. This positive feedback loop poses a risk of thermal runaway instability in ITER. Consequently, radiation and transport processes between different plasma regions with varying timescales are crucial for effective burning plasma operation.

In previous research by Liu and Stacey \cite{stacey2021nodal,liu2021multi,liu2022multi,liu2024application}, a multi-region multi-timescale burning plasma dynamics model has been developed to simulate these complex interactions in tokamaks. Different regions, including the core, edge, scrape-off layer (SOL), and divertor, are modeled as separate nodes. This model incorporates essential mechanisms such as auxiliary heating, fusion alpha heating, radiations, collisional energy transfer, transport, and ion orbit loss (IOL). Neural ordinary differential equations (Neural ODEs) \cite{chen2018neural} have been employed to optimize the parametric diffusivity formula, and this multinodal model has been validated for deuterium plasmas from DIII-D \cite{liu2024application}.

In this study, we extend this multinodal burning plasma dynamics model, NeuralPlasmaODE\footnote{Repository: \url{https://github.com/zefang-liu/NeuralPlasmaODE}}, to analyze ITER D-T plasmas. We make several assumptions suitable for ITER burning plasmas, modeling deuterons, tritons, alpha particles, and electrons in the core and edge nodes. This specialized multinodal model is applied to simulate inductive and non-inductive operational scenarios, examining the dynamics and energy flows among species through various mechanisms with multiple timescales. By leveraging Neural ODEs for the numerical derivation of diffusivity parameters, we enhance the model's precision and efficiency. Additionally, we employ transfer learning to utilize model parameters derived from fitting DIII-D data and avoid the need for training from scratch. The simulation results indicate that radiation and transport processes efficiently remove excess heat from fusion alpha particles, preventing thermal runaway instability. This study highlights the potential of machine learning in improving our understanding and control of burning plasma dynamics in fusion reactors.

%%%%%%%%%%%%%%%%%%%%%%%%%%%%%%%%%%%%%%%%
\section{Related Work}

Research on fusion-relevant plasmas, particularly focusing on burning plasma dynamics, has been extensive. \citet{wang1997simulation} used a 1.5D Tokamak Transport Simulation Code (TTSC) to simulate ITER burning plasmas, highlighting potential issues with fusion power excursions due to slight confinement improvements. \citet{green2003iter} categorized burning plasma physics in ITER into energetic particle effects, self-heating phenomena, and reactor-scale physics, assessing various drive scenarios. \citet{cordey2005scaling} improved the ELMy H-mode scaling law for energy confinement time using principal component regression, enhancing ITER performance predictions. \citet{stacey2007survey} reviewed abrupt transition phenomena in plasmas and studied theoretical thermal instabilities. \citet{hill2017confinement} developed a confinement tuning model specific to DIII-D experiments, showing improved temperature simulations over the ITER-98 scaling law. \citet{hill2019burn} further investigated control mechanisms for plasma power excursion in ITER, identifying electron cyclotron radiation (ECR) as a critical passive control mechanism, and presented a framework for multinodal dynamics modeling. \citet{stacey2021nodal} introduced a spatially coarse nodal space-time dynamics model for burning plasmas, outlining core and edge particle and energy balance equations. \citet{liu2024application} proposed a multi-region multi-timescale transport model employing neural ordinary differential equations (Neural ODEs) to simulate intricate energy transfer processes in tokamaks, validated against DIII-D experimental data. These studies lay the foundation for our work, which builds a practical multinodal model using Neural ODEs and machine learning to simulate ITER plasmas.

%%%%%%%%%%%%%%%%%%%%%%%%%%%%%%%%%%%%%%%%
\section{Burning Plasma Dynamics Model}

To effectively and efficiently simulate the complex energy transfer processes in ITER burning deuterium-tritium (D-T) plasmas, we implement a multi-region multi-timescale burning plasma dynamics model.

%%%%%%%%%%%%%%%%%%%%%%%%%%%%%%%%%%%%%%%%
\subsection{ITER Plasma Geometry}

\begin{figure}[!h]
    \centering
    \begin{subfigure}{0.6\linewidth}
        \centering
        \includegraphics[width=\textwidth]{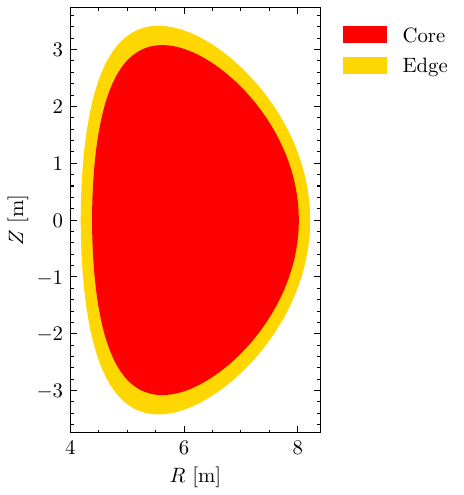}
        \caption{Cross section of an ITER plasma}
    \end{subfigure}
    \vskip 12pt
    \begin{subfigure}{0.6\linewidth}
        \centering
        \includegraphics[width=\textwidth]{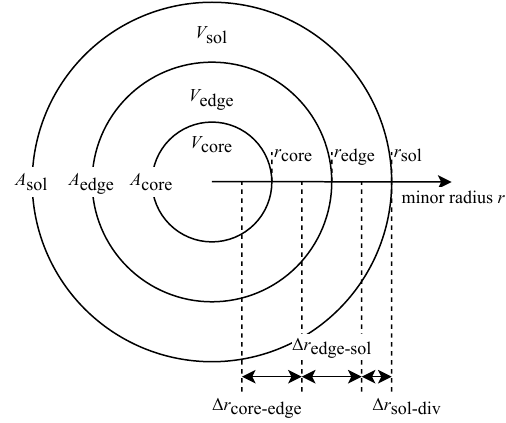}
        \caption{Geometry of the multinodal model}
    \end{subfigure}
    \caption{Multinodal model geometry of tokamak plasmas, where the first figure shows the cross section of an ITER plasma, and the second figure is the simplified geometry in the multinodal model.}
    \label{fig:iter-regions}
\end{figure}

The geometry of the burning plasma dynamics model is fundamental for simulating ITER plasmas. A conventional tokamak is viewed as a torus with a circular cross-section. The torus is divided into three regions: the core, edge, and scrape-off layer (SOL), following the flux surfaces from the inner to the outer side, while the divertor region is ignored in this geometry. Each region is represented as a separate node in this model, as illustrated in Figure \ref{fig:iter-regions}. In this model, each node is a toroidal shell with interfaces represented as torus surfaces. The minor radii for the surfaces $r_{\C}$, $r_{\E}$, and $r_{\SOL}$ correspond to the core, edge, and SOL surfaces respectively. The radial distances between these nodes are defined as $\Delta r_{\CE}$ for the core to edge, $\Delta r_{\ES}$ for the edge to SOL, and $\Delta r_{\SD}$ for the distance from the SOL node center to its outer surface. These radial distances are calculated as $\Delta r_{\CE} = r_{\E} / 2$, $\Delta r_{\ES} = ( r_{\SOL} - r_{\C}) / 2$, and $\Delta r_{\SD} = (r_{\SOL} - r_{\E}) / 2$. Additionally, the normalized minor radius is defined as $\rho = r / a$, where $a$ is the minor radius of the plasma.

%%%%%%%%%%%%%%%%%%%%%%%%%%%%%%
\subsection{Assumptions for ITER Plasmas}

To simulate ITER D-T plasmas, the multinodal burning plasma model \cite{liu2024application} requires several necessary modifications to balance accuracy and computational efficiency:
\begin{itemize}
    \item Only the core and edge regions are modeled as separate nodes.
    \item The model computes the behavior of deuterons, tritons, alpha particles, and electrons, while including helium, beryllium, and argon as impurity particles.
    \item The simulation focuses solely on the D-T fusion reaction, while all atomic and molecular processes, as well as neutral and recycling particles, are neglected.
    \item The triton particle and thermal diffusivities are assumed to be equal to those of deuterons.
    \item A delay mechanism for fusion alpha heating is implemented in the core node. 
    \item Predetermined energy deposition profiles for neutral beam injection (NBI) and radiofrequency (RF) heating \cite{kessel2007simulation} are utilized.
    \item Electron cyclotron radiation (ECR) \cite{albajar2001improved,albajar2009raytec} parameters, fitted for typical ITER profiles, are applied.
    \item Ion orbit loss (IOL) \cite{stacey2011effect,stacey2013effect,wilks2016improvements} is considered only for the edge node, with approximated loss timescales equated to transport times.
\end{itemize}

%%%%%%%%%%%%%%%%%%%%%%%%%%%%%%%%%%%%%%%%
\subsection{Particle and Energy Balance Equations}
\label{sec:balance-equations}

Based on the aforementioned assumptions, the multinodal burning plasma model is presented into the following particle and energy balance equations.

%%%%%%%%%%%%%%%%%%%%%%%%%%%%%%%%%%%%%%%%
\subsubsection{Particle Balance Equations}

Particle balance equations for deuterons ($\ce{D}$), tritons ($\ce{T}$), and alpha particles ($\alpha$) in the core and edge nodes are
\begin{align}
    \frac{\dif n_{\sigma}^{\C}}{\dif t} &= S_{\sigma,\text{ext}}^{\C} + S_{\sigma,\text{fus}}^{\C} + S_{\sigma\text{tran}}^{\C} , \\
    \frac{\dif n_{\sigma}^{\E}}{\dif t} &= S_{\sigma,\text{ext}}^{\E} + S_{\sigma,\text{fus}}^{\E} + S_{\sigma,\text{tran}}^{\E} + S_{\sigma,\IOL}^{\E} , 
\end{align}
where $ \sigma \in \set{\ce{D}, \ce{T}, \alpha} $. The electron densities are computed from the charge neutrality:
\begin{align}
    n_e^{\R} &= z_{\ce{D}} n_{\ce{D}}^{\R} + z_{\ce{T}} n_{\ce{T}}^{\R} + z_{\alpha} n_{\alpha}^{\R} + \sum_{z} z_{z} n_{z}^{\R} , 
\end{align}
where $ \R \in \set{\C, \E} $ and $z$ is for impurity particles. The particle source and sink terms are as follows.

The \textbf{external particle sources} are computed by summing the neutral beam injection (NBI), gas puffing (GAS), shatter pellet injection (SPI), and other external particle sources:
\begin{equation}
    S_{\sigma, \text{ext}}^{\R} = S_{\sigma, \NBI}^{\R} + S_{\sigma, \GAS}^{\R} + S_{\sigma, \SPI}^{\R}  + \dots ,
\end{equation}
where $ \sigma \in \set{\ce{D}, \ce{T}, \alpha} $. In this research, only particle sources from the NBI \cite{wesson2011tokamaks} are considered. The \textbf{fusion terms} are computed from the D-T fusion reactions \cite{stacey2012fusion} by
\begin{align}
    S_{\ce{D}, \text{fus}}^{\R} &= S_{\ce{T}, \text{fus}}^{\R} = - n_{\ce{D}}^{\R} n_{\ce{T}}^{\R} \braket{\sigma v}_{\text{fus}} , \\
    S_{\alpha, \text{fus}}^{\R} &= n_{\ce{D}}^{\R} n_{\ce{T}}^{\R} \braket{\sigma v}_{\text{fus}} ,
\end{align}
where $ \braket{\sigma v}_{\text{fus}} $ is the fusion reactivity \cite{bosch1992improved}. The \textbf{particle transport terms} in the core and edge nodes are computed by
\begin{align}
    S_{\sigma,\text{tran}}^{\C} &= - \frac{n_{\sigma}^{\C} - n_{\sigma}^{\E}}{\tau_{P, \sigma}^{\C \to \E}} , \\
    S_{\sigma, \text{tran}}^{\E} &= \frac{V_{\C}}{V_{\E}} \frac{n_{\sigma}^{\C} - n_{\sigma}^{\E}}{\tau_{P, \sigma}^{\C \to \E}} - \frac{n_{\sigma}^{\E}}{\tau_{P, \sigma}^{\E \to \SOL}} ,
\end{align}
where $ \tau_{P,\sigma}^{\C \to \E} $ (or $ \tau_{P,\sigma}^{\E \to \SOL} $) is the particle transport time from the core (or edge) node to the edge (or SOL) node. These particle transport times \cite{liu2022multi} are calculated by
\begin{align}
    \tau_{P, \sigma}^{\C \to \E} &= \frac{r_{\C}^2}{2 r_{\C}} \frac{\Delta r_{\CE}}{D_{\sigma}^{\C}} , \\
    \tau_{P, \sigma}^{\E \to \SOL} &= \frac{r_{\E}^2 - r_{\C}^2}{2 r_{\E}} \frac{\Delta r_{\ES}}{D_{\sigma}^{\E}} ,
\end{align}
where $ D_{\sigma}^{\C} $ and $ D_{\sigma}^{\E} $ are the core and edge particle diffusivities respectively. The \textbf{ion orbit loss (IOL) terms} are computed by $ S_{\sigma, \IOL}^{\E} = - F_{\sigma, \text{orb}}^{\E}/ \tau_{P, \sigma, \IOL}^{\E} \cdot n_{\sigma}^{\E} $, where $ F_{\sigma, \text{orb}}^{\E} $ is the particle loss fraction \cite{stacey2011effect,stacey2015distribution}, and $ \tau_{P, \sigma, \IOL}^{\E} $ is the particle IOL timescale.

%%%%%%%%%%%%%%%%%%%%%%%%%%%%%%%%%%%%%%%%
\subsubsection{Energy Balance Equations}

Energy balance equations for deuterons, tritons, alpha particles, and electrons in the core and edge nodes are
\begin{align}
    \frac{\dif U_{\sigma}^{\C}}{\dif t} &= P_{\sigma \text{aux}}^{\C} + P_{\sigma,\text{fus}}^{\C} + Q_{\sigma}^{\C} + P_{\sigma,\text{tran}}^{\C} , \\
    \frac{\dif U_{\sigma}^{\E}}{\dif t} &= P_{\sigma, \text{aux}}^{\E} + P_{\sigma,\text{fus}}^{\E} +Q_{\sigma}^{\E} + P_{\sigma,\text{tran}}^{\E} + P_{\sigma,\IOL}^{\E} , \\
    \begin{split}
        \frac{\dif U_{e}^{\C}}{\dif t} &= P_{\Omega}^{\C} + P_{e, \text{aux}}^{\C} + P_{e,\text{fus}}^{\C} - P_{R}^{\C} + Q_{e}^{\C} \\
        & \quad + P_{e,\text{tran}}^{\C} ,
    \end{split} \\
    \begin{split}
        \frac{\dif U_{e}^{\E}}{\dif t} &= P_{\Omega}^{\E} + P_{e, \text{aux}}^{\E} + P_{e,\text{fus}}^{\E} - P_{R}^{\C} + Q_{e}^{\E} \\
        & \quad + P_{e,\text{tran}}^{\E} , 
    \end{split}
\end{align}
where $ \sigma \in \set{\ce{D}, \ce{T}, \alpha} $,  the nodal energy density is defined as $ U_{\sigma}^{\R} = 3 / 2 \cdot n_{\sigma}^{\R} T_{\sigma}^{\R} $. The energy source and sink terms are as follows.

The \textbf{ohmic heating power} \cite{stacey2012fusion} is computed from the plasma current $ I_P $ by
\begin{equation}
    P_{\Omega}^{\R} \left( \si{W/m^3} \right) = \SI{2.8E-9}{} \frac{Z_{\text{eff}} I_P^2}{a^4 T_e^{3/2}} ,
\end{equation}
where $ Z_{\text{eff}} $ is the effective atomic number, the plasma current $ I_P $ is in \si{A}, the minor radius $ a $ in \si{m}, and the nodal electron temperature $ T_e = T_e^{\R} $ in \si{keV}. The \textbf{auxiliary heating terms} contain the neutral beam injection (NBI) and radiofrequency (RF) heating (including ion cyclotron heating (ICH), electron cyclotron heating (ECH), and lower hybrid heating (LH)):
\begin{align}
    P_{i, \text{aux}}^{\R} &= P_{i, \NBI}^{\R} + P_{i, \ICH}^{\R} , \\
    P_{e, \text{aux}}^{\R} &= P_{e, \NBI}^{\R} + P_{e, \ECH}^{\R} + P_{\sigma, \LH}^{\R} .
\end{align}
The \textbf{fusion power} is
\begin{equation}
    P_{\sigma, \text{fus}}^{\R} = n_{\ce{D}}^{\R} n_{\ce{T}}^{\R} \braket{\sigma v}_{\text{fus}} U_{f\sigma} ,
\end{equation}
where $ U_{f\sigma} $ is the fusion energy transferred to the species $ \sigma $ calculated by NBI heating formulas \cite{wesson2011tokamaks}. The delay effect between fusion alpha heating to electrons and ions is considered by a slowing-down timescale $ \tau_{se}^{\R} $ as $ \tau_{\text{se}} = (3 \sqrt{2 \pi} T_e^{3/2}) / (m_e^{1/2} m_b A_D) $. The fusion reaction rate $ S_{\text{fus}}^{\R} $ for core ions is evaluated at
\begin{equation}
    T_i^{\R} (t - \tau_{se}) \approx T_i^{\R} (t) - \frac{\dif T_i^{\R}}{\dif t} \bigg|_{t} \cdot \tau_{se} ,
\end{equation}
so the fusion heating to ions is postponed by this timescale compared with electrons. \textbf{Collisional energies} transferred between ions and electrons are
\begin{align}
    Q_{\alpha}^{\R} &= Q_{\alpha \ce{D}}^{\R} + Q_{\alpha \ce{T}}^{\R} + Q_{\alpha e}^{\R} , \\
    Q_{e}^{\R} &= - Q_{\alpha e}^{\R} - Q_{\ce{D} e}^{\R} - Q_{\ce{T} e}^{\R} , \\
    Q_{\ce{D}}^{\R} &= - Q_{\alpha \ce{D}}^{\R} + Q_{\ce{D} \ce{T}}^{\R} + Q_{\ce{D} e}^{\R} , \\
    Q_{\ce{T}}^{\R} &= - Q_{\alpha \ce{T}}^{\R} - Q_{\ce{D} \ce{T}}^{\R} + Q_{\ce{T} e}^{\R} ,
\end{align}
where $ Q_{\sigma \sigma'}^{\R} $ is the energy transferred from the species $ \sigma' $ to $ \sigma $ \cite{stacey2012fusion}. The \textbf{radiation terms} from electron cyclotron radiation (ECR) \cite{albajar2001improved}, bremsstrahlung \cite{stacey2012fusion}, and impurity radiation \cite{stacey2012fusion,roberts1981total,morozov2007impurity} are computed by
\begin{equation}
    P_R^{\R} = P_{\ECR}^{\R} + P_{\text{brem}}^{\R} + P_{\text{imp}}^{\R} .
\end{equation}
The \textbf{energy transport terms} in the core and edge nodes are
\begin{align}
    P_{\sigma,\text{tran}}^{\C} &= - \frac{U_{\sigma}^{\C} - U_{\sigma}^{\E}}{\tau_{E, \sigma}^{\C \to \E}} , \\
    P_{\sigma, \text{tran}}^{\E} &= \frac{V_{\C}}{V_{\E}} \frac{U_{\sigma}^{\C} - U_{\sigma}^{\E}}{\tau_{E, \sigma}^{\C \to \E}} - \frac{U_{\sigma}^{\E}}{\tau_{E, \sigma}^{\E \to \SOL}} , 
\end{align}
where $ \tau_{E,\sigma}^{\C \to \E} $ (or $ \tau_{E,\sigma}^{\E \to \SOL} $) is the energy transport time from the core (or edge) node to the edge (or SOL) node. These energy transport times \cite{liu2022multi} are solved from
\begin{align}
    \tau_{E, \sigma}^{\C \to \E} &= \frac{r_{\C}^2}{2 r_{\C}} \frac{\Delta r_{\CE}}{\chi_{\sigma}^{\C}} , \\
    \tau_{E, \sigma}^{\E \to \SOL} &= \frac{r_{\E}^2 - r_{\C}^2}{2 r_{\E}} \frac{\Delta r_{\ES}}{\chi_{\sigma}^{\E}} ,
\end{align}
where $ \chi_{\sigma}^{\C} $ and $ \chi_{\sigma}^{\E} $ are the core and edge thermal diffusivities respectively. The \textbf{IOL terms} for the edge node are computed by $ P_{\ce{D},\IOL}^{\E} = - E_{\ce{D}, \text{orb}}^{\E} / \tau_{E, \ce{D}, \IOL}^{\E} \cdot U_{\ce{D}}^{\E} $, where $ E_{\ce{D}, \text{orb}}^{\E} $ is the energy loss fraction \cite{stacey2011effect,stacey2015distribution}, and $ \tau_{E, \ce{D}, \IOL}^{\E} $ is the energy IOL timescale.

%%%%%%%%%%%%%%%%%%%%%%%%%%%%%%%%%%%%%%%%
\subsection{Diffusivity Models}

To compute internodal transport times, it is essential to have formulas for particle and thermal diffusivities. An empirical scaling for the effective thermal diffusivity in ELMy H-mode tokamak plasmas \cite{becker2004study} is given by:
\begin{equation}
\begin{split}
    \chi_{H98}(\rho) & = \alpha_{H} B_T^{-3.5} n_e(\rho)^{0.9} T_e(\rho) \abs{\nabla T_e(\rho)}^{1.2} \\
    & \cdot q(\rho)^{3.0} \kappa(\rho)^{-2.9} M^{-0.6} R^{0.7} a^{-0.2} \left( \si{m^{2} / s} \right) , \label{eqn:chi-h98}
\end{split}
\end{equation}
where the thermal diffusivity $ \chi_{H98} $ in $ \si{m^2/s} $, normalized radius $ \rho = r / a $, coefficient $ \alpha_{H} = 0.123 $, toroidal magnetic field $ B_T $ in $ \si{T} $, electron density $ n_e $ in $ \SI{E19}{m^{-3}} $, electron temperature $ T_e $ in $  \si{keV} $, electron temperature gradient $ \nabla T_e $ in $ \si{keV/m} $, safety factor $ q = q_{\psi} $, local elongation $ \kappa $, hydrogenic atomic mass number $ M $ in $ \SI{1}{amu} $, major radius $ R $ in $ \si{m} $, and minor radius $ a $ in $ \si{m} $. The particle and thermal diffusivities for electrons and ions \cite{becker2006anomalous} can be assumed as $ \chi_e (\rho) = \chi_i(\rho) = \chi_{H98}(\rho) $ and $ D_i(\rho) = 0.6 \chi_{H98}(\rho) $. This empirical scaling was used as the baseline by \citet{liu2024application}.

For modeling ITER plasmas accurately, we apply a parametric diffusivity formula for diffusivities \cite{liu2024application}:
\begin{equation}
\begin{split}
    & \frac{\chi(\rho)}{\SI{1}{m^2/s}} = \alpha_H \left( \frac{B_T}{\SI{1}{T}} \right)^{\alpha_{B}} \left( \frac{n_e(\rho)}{\SI{E19}{m^{-3}}} \right)^{\alpha_n} \\
    & \cdot \left( \frac{T_e(\rho)}{\SI{1}{keV}} \right)^{\alpha_T} \left( \frac{\abs{\nabla T_e(\rho)}}{\SI{1}{keV/m}} \right)^{\alpha_{\nabla T}} q(\rho)^{\alpha_q}  \\
    & \cdot \kappa(\rho)^{\alpha_{\kappa}} \left( \frac{M}{\SI{1}{amu}} \right)^{\alpha_M} \left( \frac{R}{\SI{1}{m}} \right)^{\alpha_{R}} \left( \frac{a}{\SI{1}{m}} \right)^{\alpha_{a}} .
\end{split} \label{eqn:chi-scaling}
\end{equation}
where $ \alpha_H, \alpha_B, \dots,  \alpha_a $ are parameters to be determined. This diffusivity formula can be expressed in vector form as $ \ln \bm{\chi}_{\R} = \mathbf{b}_{\R} + \mathbf{W}_{\R} \ln \mathbf{x}_{\R} $, where $ \bm{\chi}_{\R} $ represents the internodal diffusivities vector, $ \mathbf{b}_{\R} $ is the bias vector, $ \mathbf{W}_{\R} $ is the weight matrix, and $ \mathbf{x}_{\R} $ is the vector of corresponding physical values.

%%%%%%%%%%%%%%%%%%%%%%%%%%%%%%%%%%%%%%%%
\section{Computational Framework}

\begin{figure}[!h]
\centering
\includegraphics[width=\linewidth]{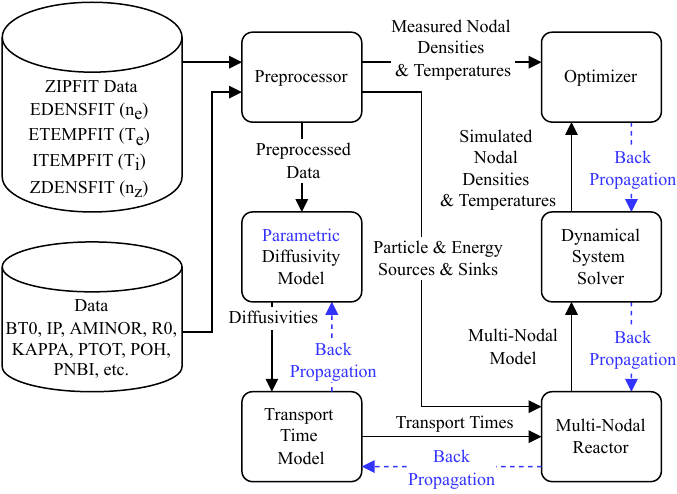}
\caption{Computational framework of NeuralPlasmaODE, including cylinders as datasets, squares as modules, solid lines as forward flows, and dashed lines as back propagation processes.}
\label{fig:framework}
\end{figure}

In this study, we enhance the computational framework of NeuralPlasmaODE \cite{liu2024application} to simulate burning plasma dynamics in ITER. The framework consists of several modules designed for processing experimental data, simulating plasma behavior, and optimizing model parameters. Figure \ref{fig:framework} shows a workflow diagram of this framework.

Initially, a data module reads inputs such as two-dimensional plasma profiles and one-dimensional global parameters. A preprocessing module then standardizes these inputs into uniform time sequences and performs volume averaging on two-dimensional signals to derive nodal particle densities and temperatures. Central to the framework is a diffusivity model that calculates particle and thermal diffusivities based on experimental conditions, along with a transport time model that determines the timescales for particle and energy transport between nodes. These models feed into a reactor simulation module that integrates sources, sinks, and transport terms into a dynamical system. This system is solved using the neural ordinary differential equation (Neural ODE) \cite{chen2018neural} solver, which outputs estimated particle densities and temperatures. These estimates are then refined through an optimization module, which calculates the mean square error (MSE) between model predictions and experimental data. Using back-propagation for gradient computation and gradient descent for parameter updates, this module ensures the model parameters are optimized for accurate plasma behavior representation.

%%%%%%%%%%%%%%%%%%%%%%%%%%%%%%%%%%%%%%%%
\section{Simulations for ITER Plasmas}

In this section, we present our simulations for ITER plasmas, including introducing the various simulation scenarios, discussing the simulation methodology, and analyzing the simulation results.

%%%%%%%%%%%%%%%%%%%%%%%%%%%%%%%%%%%%%%%%
\subsection{Simulation Scenarios}

\begin{figure}[!h]
    \centering
    \begin{subfigure}{0.45\linewidth}
        \centering
        \includegraphics[width=\textwidth]{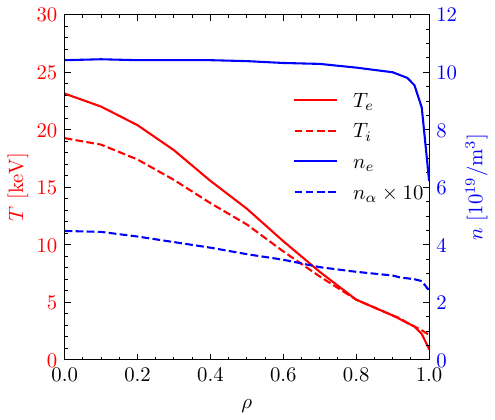}
        \caption{Inductive operation scenario}
    \end{subfigure}
    \hfill
    \begin{subfigure}{0.45\linewidth}
        \centering
        \includegraphics[width=\textwidth]{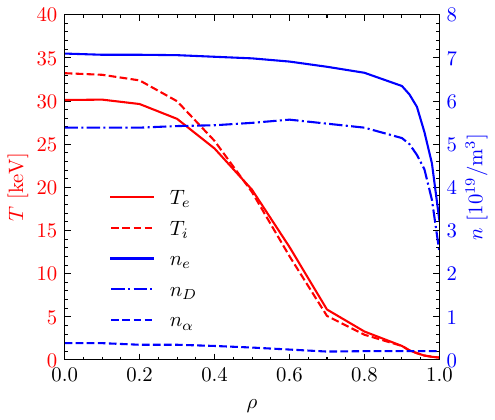}
        \caption{Non-inductive operation scenario}
    \end{subfigure}
    \caption{Typical radial profiles of plasma temperatures and densities in the ITER inductive and non-inductive operation scenarios (reproduced with permission from \citet{iaea2002iter}).}
    \label{fig:iter-profiles}
\end{figure}

The multinodal burning plasma model (NeuralPlasmaODE) is applied to simulate the ITER deuterium-tritium (D-T) plasma. The ITER heating and current drive (H\&CD) system \cite{iaea2002iter,shimada2007overview,henderson2015targeted} includes 33 MW neutral beam injection (NB), 20 MW ion cyclotron heating (IC), and 20 MW electron cyclotron heating (EC) in the initial campaign, with potential upgrades to 50 MW NB, 40 MW IC, 40 MW EC, and 40 MW lower hybrid heating (LH) in the future. The ITER tokamak can operate in both inductive and non-inductive modes. In the inductive mode, the ohmic current is the primary contributor to the total toroidal current, while in the non-inductive mode, highly energetic neutral atom injection and powerful radiofrequency radiation drive most of the toroidal current. For this study, we select inductive scenario 2, hybrid scenario 3, and non-inductive scenario 4 from the ITER design \cite{iaea2002iter}. Their typical operating conditions are shown in Table \ref{tab:iter-parameters}. The important parameters are contained, including geometries, electromagnetic values, auxiliary heating powers, densities, temperatures, and impurity fractions. Additionally, the typical radial profiles of electron and ion temperatures, and electron and helium densities are depicted in Figure \ref{fig:iter-profiles}. These profiles are integrated over the core and edge nodes to obtain nodal densities and temperatures, which are used to adjust the diffusivity parameters.

\begin{table*}[!h]
    \centering
    \small
    \begin{tabular}{lllll}
        \toprule
        & & \textbf{Inductive} & \textbf{Hybrid} & \textbf{Non-Inductive} \\
        \textbf{Parameter} & \textbf{Symbol} & \textbf{Scenario 2} & \textbf{Scenario 3} & \textbf{Scenario 4} \\
        \midrule
        Major radius & $ R_0 $ (\SI{}{m}) & 6.2 & 6.2 & 6.35 \\
        Minor radius & $ a $ (\SI{}{m}) & 2.0 & 2.0 & 1.85 \\
        Volume & $ V $ ($ \SI{}{m^3} $) & 831 & 831 & 794 \\
        Surface & $ S $ ($ \SI{}{m^2} $) & 683 & 683 & - \\
        Elongation at the 95\% flux surface & $ \kappa_{95} $ & 1.70 & 1.70 & 1.85 \\
        Triangularity at the 95\% flux surface & $ \delta_{95} $ & 0.33 & 0.33 & 0.40 \\
        \midrule
        Toroidal magnetic field at the magnetic axis & $ B_T $ (\SI{}{T}) & 5.3 & 5.3 & 5.18 \\
        Plasma current & $ I_P $ (\SI{}{MA}) & 15 & 13.8 & 9.0 \\
        Safety factor at the 95\% flux surface & $ q_{95} $ & 3.0 & 3.3 & 5.3 \\
        \midrule
        Volume-averaged electron density & $ \braket{n_e} $ ($ \SI{E19}{m^{-3}} $) & 10.1 & 9.3 & 6.7 \\
        Volume-averaged ion temperature & $ \braket{T_i} $ (\SI{}{keV}) & 8.0 & 8.4 & 12.5 \\
        Volume-averaged electron temperature & $ \braket{T_e} $ (\SI{}{keV}) & 8.8 & 9.6 & 12.3 \\
        \midrule
        Fusion power & $ P_{\text{fus}} $ (\SI{}{MW}) & 400 & 400 & 356 \\
        Auxiliary heating power & $ P_{\text{aux}} $ (\SI{}{MW}) & 40 & 73 & 59 \\
        Radiofrequency heating power & $ P_{\RF} $ (\SI{}{MW}) & 7 & 40 & - \\
        Lower hybrid heating power & $ P_{\LH} $ (\SI{}{MW}) & - & - & 29 \\
        Neutral beam heating power & $ P_{\NBI} $ (\SI{}{MW}) & 33 & 33 & 30 \\
        Fusion energy gain factor & $ Q $ & 10 & 5.4 & 6.0 \\
        Energy confinement time & $ \tau_E $ (s) & 3.7 & 2.73 & 3.1 \\
        Burn time & $ t $ (\SI{}{s}) & 400 & 1070 & 3000 \\
        \midrule
        Helium fraction & $ f_{\ce{He}} $ (\%) & 3.2 & 2.5 & 4.1 \\
        Beryllium fraction & $ f_{\ce{Be}} $ (\%) & 2.0 & 2.0 & 2.0 \\
        Argon fraction & $ f_{\ce{Ar}} $ (\%) & 0.12 & 0.19 & 0.26 \\
        Effective impurity charge & $ Z_{\text{eff}} $ & 1.66 & 1.85 & 2.07 \\
        Radiation power & $ P_{\text{rad}} $ (\SI{}{MW}) & 47 & 55 & 37.6 \\
        \bottomrule
    \end{tabular}
    \caption{Typical parameters of inductive, hybrid, and non-inductive ITER operation scenarios (reproduced with permission from \citet{iaea2002iter}).}
    \label{tab:iter-parameters}
\end{table*}

For simulating the ITER design scenarios, we model several essential particle and energy sources. For external particle sources, neutral beam injection introduces negative deuteron particles into the tokamak plasma, with an equal amount of tritons assumed to be supplied to maintain a balance between deuteron and triton particles. For external energy sources, neutral beam and radiofrequency (RF) heating systems provide auxiliary heating. In the inductively driven scenario, \citet{wagner2010heating} concluded that the exact proportions of NB, IC, and EC heating are not critical, but IC heating should be utilized to heat ions directly. Therefore, we assume all RF heating power is deposited into ions in this scenario.

%%%%%%%%%%%%%%%%%%%%%%%%%%%%%%%%%%%%%%%%
\subsection{Simulation Methodology}

To accurately model the ITER burning plasma, the parameters in the diffusivity model need to be properly tuned using machine learning. In the previous study on simulating DIII-D plasmas with NeuralPlasmaODE, \citet{liu2024application} split experimental data into training and testing datasets. The training dataset was used to tune parameters in the diffusivity model, while the testing dataset evaluated the optimized model. However, since the ITER is still under construction and its experimental data are unavailable, we adopt transfer learning \cite{weiss2016survey}, where the diffusivity parameters learned from DIII-D deuterium plasmas are transferred to ITER plasmas. We then apply a fine-tuning method to these diffusivity parameters using burning simulation results from prior research \cite{iaea2002iter,green2003iter} as optimization targets. The parameters in the diffusivity model are adjusted to match the ITER design scenarios during the current flat-top phase. Such current flat-top phase is used as the training set, while the plasma start-up phase serves as the testing set.

The optimization objective is defined as a vector including densities and temperatures in the core and edge nodes:
\begin{equation}
\begin{split}
    \left[ \frac{n_{\ce{D}}^{\R}}{\SI{E19}{m^{-3}}}, \frac{n_{\alpha}^{\R}}{\SI{E18}{m^{-3}}}, \frac{n_{e}^{\R}}{\SI{E19}{m^{-3}}}, \frac{T_{\ce{D}}^{\R}}{\SI{1}{keV}}, \frac{T_{e}^{\R}}{\SI{1}{keV}} \right] ,
\end{split}
\end{equation}
with the nodal diffusivity parameters initialized to those of the DIII-D plasma. Densities and temperatures are initialized at current flat-top values instead of a cold plasma, with a time step of 0.2 s and a total simulation time of 10 s. After 14 epochs, the mean squared error (MSE) loss for scenario 2 in the current flat-top phase drops from 6.7085 to 0.0016 with a learning rate of 0.02. The optimized multinodal model is then used to simulate both inductive and hybrid scenarios. For non-inductive scenarios, diffusivity parameters are transferred from inductive scenarios and trained on scenario 4 with 2 epochs and a learning rate of 0.02, reducing the MSE loss from 13.2741 to 0.6333. After the optimization, initial temperatures are reset to \SI{2}{keV} in the core node and \SI{1}{keV} in the edge node for all species, with particle densities from Table \ref{tab:iter-parameters} and Figure \ref{fig:iter-profiles}, except alpha particle densities set to \SI{e17}{m^{-3}}. The fine-tuned multinodal model is then simulated for ITER scenarios during the plasma start-up phase over a total time of 15 s for testing.

%%%%%%%%%%%%%%%%%%%%%%%%%%%%%%%%%%%%%%%%
\subsection{Simulation Results}

In this subsection, we show the results of our simulations for the different ITER operation scenarios. The outcomes for inductive, hybrid, and non-inductive scenarios are analyzed and discussed in detail.

%%%%%%%%%%%%%%%%%%%%%%%%%%%%%%%%%%%%%%%%
\subsubsection{Inductive Operation Scenario}

\begin{figure}[!h]
    \centering
    \begin{subfigure}{0.48\linewidth}
        \centering
        \includegraphics[width=\textwidth]{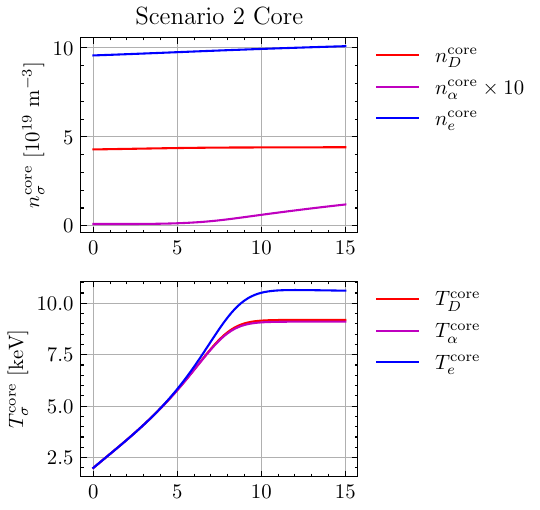}
        \caption{Core node}
    \end{subfigure}
    \hfill
    \begin{subfigure}{0.48\linewidth}
        \centering
        \includegraphics[width=\textwidth]{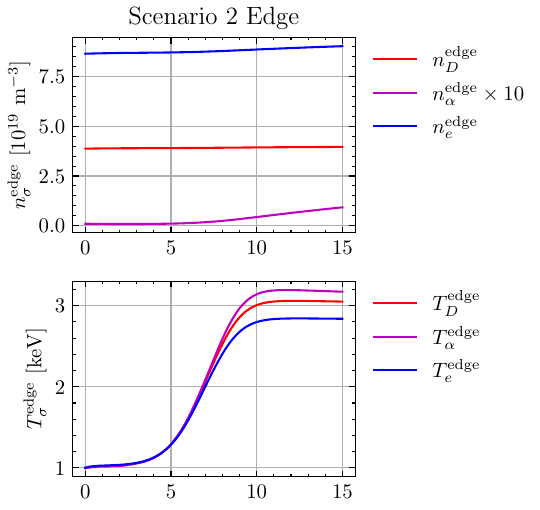}
        \caption{Edge node}
    \end{subfigure}
    \caption{Densities and temperatures of ITER inductive operation scenario 2.}
    \label{fig:iter-2-temperatures}
\end{figure}

The inductive scenario represents a mode of operation where the majority of the plasma current is driven by ohmic heating and auxiliary heating systems. The simulation results for inductive scenario 2 are presented in Figure \ref{fig:iter-2-temperatures}, showing the densities and temperatures of deuterons, alpha particles, and electrons. Triton densities and temperatures are omitted as they are approximately the same as those of deuterons. Additionally, the temperatures of alpha particles are shown only for the cold ones, excluding fusion alpha particles at \SI{3.5}{MeV} until they transfer their energy to electrons and ions. The results indicate that core and edge temperatures reach a steady state at around \SI{11}{s}, and no energy excursion due to fusion alpha heating is observed.

\begin{figure}[!h]
    \centering
    \begin{subfigure}{0.48\linewidth}
        \centering
        \includegraphics[width=\textwidth]{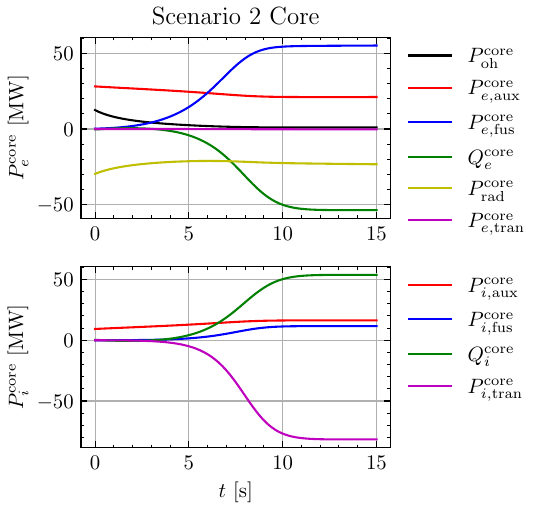}
        \caption{Core node}
    \end{subfigure}
    \hfill
    \begin{subfigure}{0.48\linewidth}
        \centering
        \includegraphics[width=\textwidth]{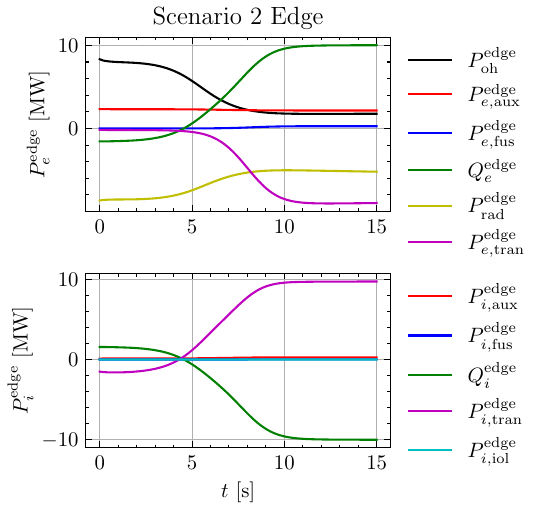}
        \caption{Edge node}
    \end{subfigure}
    \caption{Powers of ITER inductive operation scenario 2.}
    \label{fig:iter-2-powers}
\end{figure}

Figure \ref{fig:iter-2-powers} depicts the power changes over time, with positive values indicating energy gain and negative values indicating energy loss. At the start of the simulation, ohmic heating ($ P_{\text{oh}}^{\C} $) and auxiliary heating ($ P_{\sigma, \text{aux}}^{\C} $) provide most of the energy to core electrons and ions until the core ion temperature becomes high enough to initiate fusion reactions. Fusion alpha particles first heat the electrons ($ P_{e,\text{fus}}^{\C} $), after which the heated electrons transfer their energy to ions through Coulomb collisions ($ Q_{i}^{\C} = - Q_{e}^{\C} $). Meanwhile, fusion heating is also deposited directly to core ions ($ P_{i,\text{fus}}^{\C} $). The fusion heating is then removed by radiation power ($ P_{\text{rad}}^{\C} $) and transport loss ($ P_{\sigma, \text{tran}}^{\C} $).

%%%%%%%%%%%%%%%%%%%%%%%%%%%%%%
\subsubsection{Hybrid Operation Scenario}

\begin{figure}[!h]
    \centering
    \begin{subfigure}{0.48\linewidth}
        \centering
        \includegraphics[width=\textwidth]{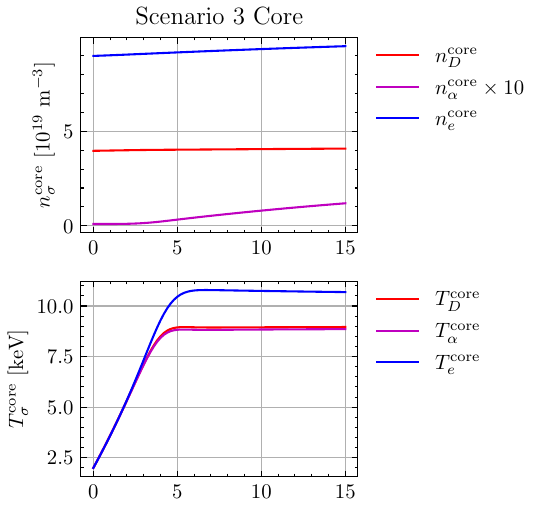}
        \caption{Core node}
    \end{subfigure}
    \hfill
    \begin{subfigure}{0.48\linewidth}
        \centering
        \includegraphics[width=\textwidth]{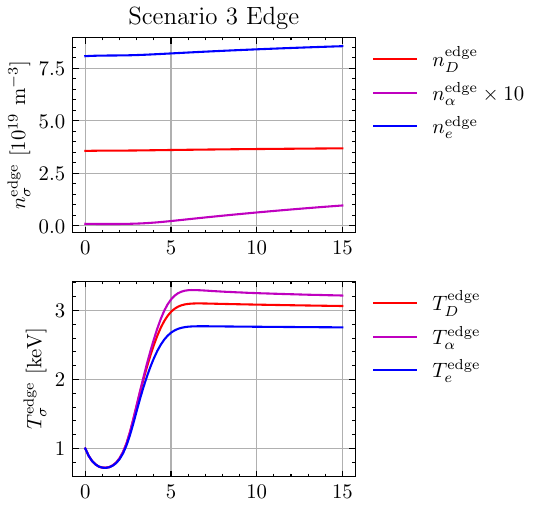}
        \caption{Edge node}
    \end{subfigure}
    \caption{Densities and temperatures of ITER hybrid operation scenario 3.}
	\label{fig:iter-3-temperatures}
\end{figure}

\begin{figure}[!h]
    \centering
    \begin{subfigure}{0.48\linewidth}
        \centering
        \includegraphics[width=\textwidth]{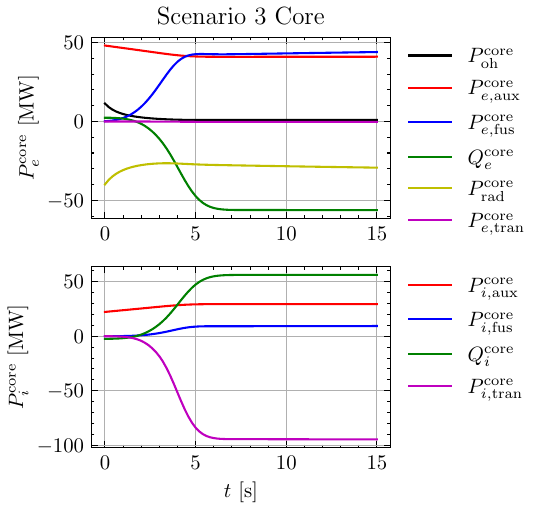}
        \caption{Core node}
    \end{subfigure}
    \hfill
    \begin{subfigure}{0.48\linewidth}
        \centering
        \includegraphics[width=\textwidth]{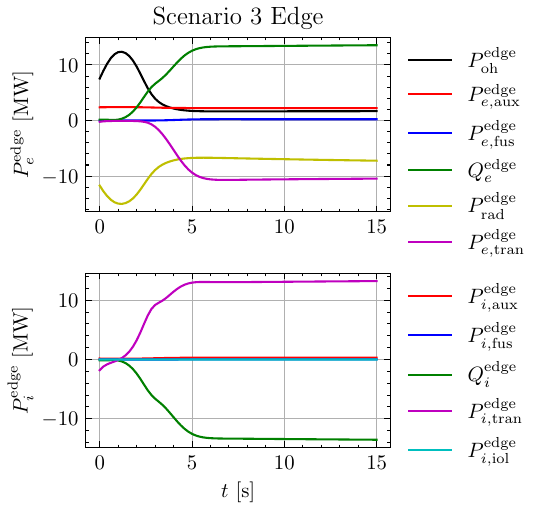}
        \caption{Edge node}
    \end{subfigure}
    \caption{Powers of ITER hybrid operation scenario 3.}
    \label{fig:iter-3-powers}
\end{figure}

A hybrid mode of operation, where a significant fraction of the plasma current is driven by non-inductive current drive power and the bootstrap current, is a promising route for establishing steady-state or non-inductive modes in ITER \cite{iaea2002iter}. To verify thermal stability with fusion alpha heating, we selected scenario 3 for simulation using the multinodal burning plasma dynamics model. The results for this scenario are shown in Figures \ref{fig:iter-3-temperatures} and \ref{fig:iter-3-powers}, first displaying the densities and temperatures, followed by the power profiles. The ion and electron temperatures reach a steady state at around \SI{6}{s}, which is shorter than in the inductive scenarios due to the higher auxiliary heating power. This increased auxiliary heating is offset by higher radiation and transport losses, preventing any power excursion in this hybrid scenario simulation.

%%%%%%%%%%%%%%%%%%%%%%%%%%%%%%%%%%%%%%%%
\subsubsection{Non-Inductive Operation Scenario}

\begin{figure}[!h]
    \centering
    \begin{subfigure}{0.48\linewidth}
        \centering
        \includegraphics[width=\textwidth]{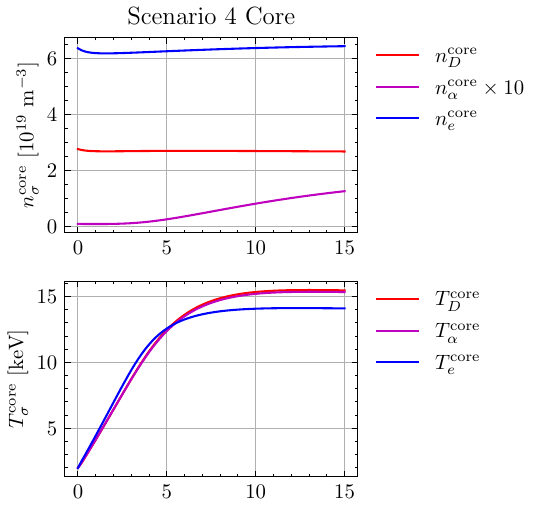}
        \caption{Core node}
    \end{subfigure}
    \hfill
    \begin{subfigure}{0.48\linewidth}
        \centering
        \includegraphics[width=\textwidth]{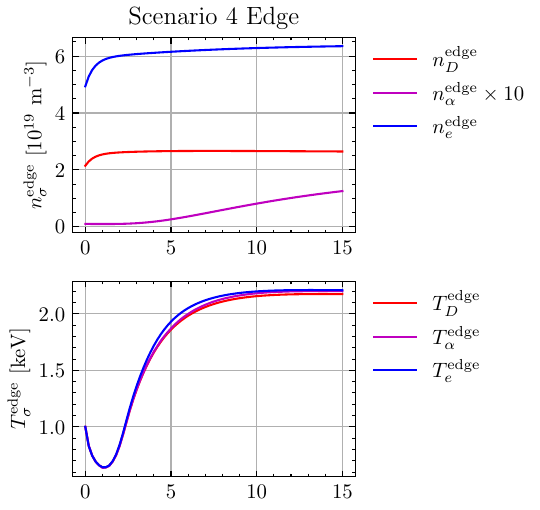}
        \caption{Edge node}
    \end{subfigure}
    \caption{Densities and temperatures of ITER non-inductive operation scenario 4.}
    \label{fig:iter-4-temperatures}
\end{figure}

\begin{figure}[!h]
    \centering
    \begin{subfigure}{0.48\linewidth}
        \centering
        \includegraphics[width=\textwidth]{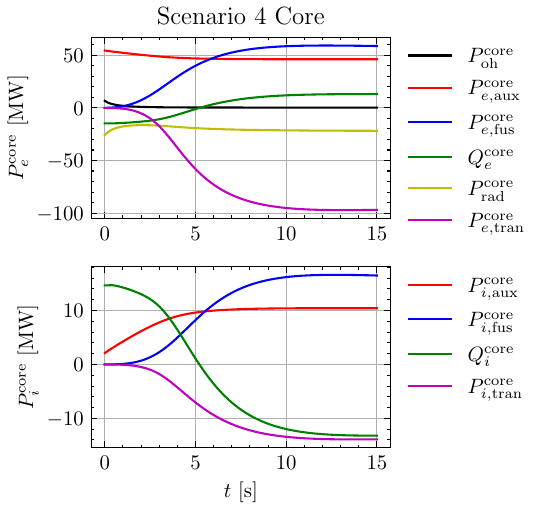}
        \caption{Core node}
    \end{subfigure}
    \hfill
    \begin{subfigure}{0.48\linewidth}
        \centering
        \includegraphics[width=\textwidth]{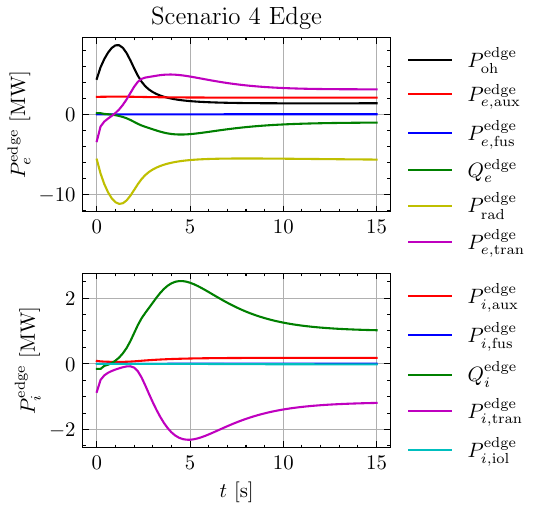}
        \caption{Edge node}
    \end{subfigure}
    \caption{Powers of ITER non-inductive operation scenario 4.}
    \label{fig:iter-4-powers}
\end{figure}

The non-inductive operation scenario selected for this study is weak negative shear (WNS) scenario 4. Compared to the inductive profiles, the typical radial profiles in Figure \ref{fig:iter-profiles} show a higher core temperature with a steeper temperature gradient. The plasma current in the non-inductive scenario is lower than in the inductive one, but the safety factor at the 95\% flux surface is higher. Additionally, the plasma current is roughly equally divided between current drive and bootstrap current. The simulation results for the non-inductive scenario 4 are presented in Figures \ref{fig:iter-4-temperatures} and \ref{fig:iter-4-powers}. The ion and electron temperatures reach a steady state at around \SI{12}{s}. Compared to inductive scenario 2, non-inductive scenario 4 exhibits higher fusion and auxiliary heating. Additionally, the energy transported from core electrons to the edge is significantly greater. However, Coulomb collisional energy transfer occurs from ions to electrons since the core electron temperature is lower than the core ion temperature. No energy excursion due to fusion alpha heating is observed in this simulation.

%%%%%%%%%%%%%%%%%%%%%%%%%%%%%%%%%%%%%%%%
\section{Conclusion}

In this research, we simulate ITER deuterium-tritium (D-T) plasmas using NeuralPlasmaODE, a multi-region multi-timescale transport model, to understand complex energy transfer processes. By employing neural ordinary differential equations (Neural ODEs) and leveraging transfer learning with parameters derived from DIII-D experimental data, we enhance simulation efficiency and accuracy. The model is applied to both inductive and non-inductive operational scenarios of ITER, showing that in inductive scenarios, core electrons dissipate energy through various radiation mechanisms, while in non-inductive scenarios, most plasma current is generated non-inductively, leading to higher core temperatures and significant energy transport to the edge. In both scenarios, radiation and transport processes effectively prevent thermal runaway instability, highlighting the potential of machine learning to advance our understanding and control of burning plasma dynamics in fusion reactors.

%%%%%%%%%%%%%%%%%%%%%%%%%%%%%%%%%%%%%%%%
\newpage
\bibliography{aaai25}

\begin{thebibliography}{30}
\providecommand{\natexlab}[1]{#1}

\bibitem[{Albajar, Bornatici, and Engelmann(2009)}]{albajar2009raytec}
Albajar, F.; Bornatici, M.; and Engelmann, F. 2009.
\newblock RAYTEC: a new code for electron cyclotron radiative transport modelling of fusion plasmas.
\newblock \emph{Nuclear Fusion}, 49(11): 115017.

\bibitem[{Albajar, Johner, and Granata(2001)}]{albajar2001improved}
Albajar, F.; Johner, J.; and Granata, G. 2001.
\newblock Improved calculation of synchrotron radiation losses in realistic tokamak plasmas.
\newblock \emph{Nuclear fusion}, 41(6): 665.

\bibitem[{Becker(2004)}]{becker2004study}
Becker, G. 2004.
\newblock Study of anomalous inward drift in tokamaks by transport analysis and simulations.
\newblock \emph{Nuclear fusion}, 44(9): 933.

\bibitem[{Becker and Kardaun(2006)}]{becker2006anomalous}
Becker, G.; and Kardaun, O. 2006.
\newblock Anomalous particle pinch and scaling of vin/D based on transport analysis and multiple regression.
\newblock \emph{Nuclear fusion}, 47(1): 33.

\bibitem[{Bosch and Hale(1992)}]{bosch1992improved}
Bosch, H.-S.; and Hale, G.~M. 1992.
\newblock Improved formulas for fusion cross-sections and thermal reactivities.
\newblock \emph{Nuclear fusion}, 32(4): 611.

\bibitem[{Chen et~al.(2018)Chen, Rubanova, Bettencourt, and Duvenaud}]{chen2018neural}
Chen, R.~T.; Rubanova, Y.; Bettencourt, J.; and Duvenaud, D.~K. 2018.
\newblock Neural ordinary differential equations.
\newblock \emph{Advances in neural information processing systems}, 31.

\bibitem[{Cordey et~al.(2005)Cordey, Thomsen, Chudnovskiy, Kardaun, Takizuka, Snipes, Greenwald, Sugiyama, Ryter, Kus et~al.}]{cordey2005scaling}
Cordey, J.; Thomsen, K.; Chudnovskiy, A.; Kardaun, O.; Takizuka, T.; Snipes, J.; Greenwald, M.; Sugiyama, L.; Ryter, F.; Kus, A.; et~al. 2005.
\newblock Scaling of the energy confinement time with $\beta$ and collisionality approaching ITER conditions.
\newblock \emph{Nuclear fusion}, 45(9): 1078.

\bibitem[{Green et~al.(2003)}]{green2003iter}
Green, B.; et~al. 2003.
\newblock ITER: burning plasma physics experiment.
\newblock \emph{Plasma physics and controlled fusion}, 45(5): 687.

\bibitem[{Henderson et~al.(2015)Henderson, Saibene, Darbos, Farina, Figini, Gagliardi, Gandini, Gassmann, Hanson, Loarte et~al.}]{henderson2015targeted}
Henderson, M.; Saibene, G.; Darbos, C.; Farina, D.; Figini, L.; Gagliardi, M.; Gandini, F.; Gassmann, T.; Hanson, G.; Loarte, A.; et~al. 2015.
\newblock The targeted heating and current drive applications for the ITER electron cyclotron system.
\newblock \emph{Physics of plasmas}, 22(2).

\bibitem[{Hill(2019)}]{hill2019burn}
Hill, M.~D. 2019.
\newblock \emph{Burn Control Mechanisms in Tokamak Fusion Reactors}.
\newblock Ph.D. thesis, Georgia Institute of Technology.

\bibitem[{Hill and Stacey(2017)}]{hill2017confinement}
Hill, M.~D.; and Stacey, W.~M. 2017.
\newblock Confinement Tuning of a 0-D Plasma Dynamics Model.
\newblock \emph{Fusion Science and Technology}, 72(2): 162--175.

\bibitem[{IAEA(2002)}]{iaea2002iter}
IAEA. 2002.
\newblock \emph{ITER Technical Basis}.
\newblock Number~24 in ITER EDA Documentation Series. Vienna: International Atomic Energy Agency.

\bibitem[{Kessel et~al.(2007)Kessel, Giruzzi, Sips, Budny, Artaud, Basiuk, Imbeaux, Joffrin, Schneider, Murakami et~al.}]{kessel2007simulation}
Kessel, C.; Giruzzi, G.; Sips, A.; Budny, R.; Artaud, J.; Basiuk, V.; Imbeaux, F.; Joffrin, E.; Schneider, M.; Murakami, M.; et~al. 2007.
\newblock Simulation of the hybrid and steady state advanced operating modes in ITER.
\newblock \emph{Nuclear Fusion}, 47(9): 1274.

\bibitem[{Liu(2022)}]{liu2022multi}
Liu, Z. 2022.
\newblock \emph{A Multi-Region Multi-Timescale Burning Plasma Dynamics Model for Tokamaks}.
\newblock Ph.D. thesis, Georgia Institute of Technology.

\bibitem[{Liu and Stacey(2021)}]{liu2021multi}
Liu, Z.; and Stacey, W. 2021.
\newblock A Multi-Region Multi-Timescale Burning Plasma Dynamics Model for Tokamaks.
\newblock In \emph{APS Division of Plasma Physics Meeting Abstracts}, volume 2021, UP11--087.

\bibitem[{Liu and Stacey(2024)}]{liu2024application}
Liu, Z.; and Stacey, W.~M. 2024.
\newblock Application of Neural Ordinary Differential Equations for Tokamak Plasma Dynamics Analysis.
\newblock \emph{arXiv preprint arXiv:2403.01635}.

\bibitem[{Morozov, Baronova, and Senichenkov(2007)}]{morozov2007impurity}
Morozov, D.~K.; Baronova, E.; and Senichenkov, I.~Y. 2007.
\newblock Impurity radiation from a tokamak plasma.
\newblock \emph{Plasma Physics Reports}, 33: 906--922.

\bibitem[{Roberts(1981)}]{roberts1981total}
Roberts, D. 1981.
\newblock Total impurity radiation power losses from steady-state tokamak plasmas.
\newblock \emph{Nuclear Fusion}, 21(2): 215.

\bibitem[{Shimada et~al.(2007)Shimada, Campbell, Mukhovatov, Fujiwara, Kirneva, Lackner, Nagami, Pustovitov, Uckan, Wesley et~al.}]{shimada2007overview}
Shimada, M.; Campbell, D.; Mukhovatov, V.; Fujiwara, M.; Kirneva, N.; Lackner, K.; Nagami, M.; Pustovitov, V.; Uckan, N.; Wesley, J.; et~al. 2007.
\newblock Overview and summary.
\newblock \emph{Nuclear Fusion}, 47(6): S1.

\bibitem[{Stacey(2007)}]{stacey2007survey}
Stacey, W.~M. 2007.
\newblock A survey of thermal instabilities in tokamak plasmas: Theory, comparison with experiment, and predictions for future devices.
\newblock \emph{Fusion science and technology}, 52(1): 29--67.

\bibitem[{Stacey(2011)}]{stacey2011effect}
Stacey, W.~M. 2011.
\newblock The effect of ion orbit loss and X-loss on the interpretation of ion energy and particle transport in the DIII-D edge plasma.
\newblock \emph{Physics of Plasmas}, 18(10).

\bibitem[{Stacey(2012)}]{stacey2012fusion}
Stacey, W.~M. 2012.
\newblock \emph{Fusion plasma physics}.
\newblock John Wiley \& Sons.

\bibitem[{Stacey(2013)}]{stacey2013effect}
Stacey, W.~M. 2013.
\newblock Effect of ion orbit loss on distribution of particle, energy and momentum sources into the tokamak scrape-off layer.
\newblock \emph{Nuclear Fusion}, 53(6): 063011.

\bibitem[{Stacey(2021)}]{stacey2021nodal}
Stacey, W.~M. 2021.
\newblock A Nodal Model for Tokamak Burning Plasma Space-Time Dynamics.
\newblock \emph{Fusion Science and Technology}, 77(2): 109--118.

\bibitem[{Stacey and Schumann(2015)}]{stacey2015distribution}
Stacey, W.~M.; and Schumann, M.~T. 2015.
\newblock The distribution of ion orbit loss fluxes of ions and energy from the plasma edge across the last closed flux surface into the scrape-off layer.
\newblock \emph{Physics of Plasmas}, 22(4).

\bibitem[{Wagner et~al.(2010)Wagner, Becoulet, Budny, Erckmann, Farina, Giruzzi, Kamada, Kaye, Koechl, Lackner et~al.}]{wagner2010heating}
Wagner, F.; Becoulet, A.; Budny, R.; Erckmann, V.; Farina, D.; Giruzzi, G.; Kamada, Y.; Kaye, A.; Koechl, F.; Lackner, K.; et~al. 2010.
\newblock On the heating mix of ITER.
\newblock \emph{Plasma Physics and Controlled Fusion}, 52(12): 124044.

\bibitem[{Wang et~al.(1997)Wang, Amano, Ogawa, and Inoue}]{wang1997simulation}
Wang, J.-F.; Amano, T.; Ogawa, Y.; and Inoue, N. 1997.
\newblock Simulation of burning plasma dynamics in ITER.
\newblock \emph{Fusion technology}, 32(4): 590--600.

\bibitem[{Weiss, Khoshgoftaar, and Wang(2016)}]{weiss2016survey}
Weiss, K.; Khoshgoftaar, T.~M.; and Wang, D. 2016.
\newblock A survey of transfer learning.
\newblock \emph{Journal of Big data}, 3: 1--40.

\bibitem[{Wesson and Campbell(2011)}]{wesson2011tokamaks}
Wesson, J.; and Campbell, D.~J. 2011.
\newblock \emph{Tokamaks}, volume 149.
\newblock Oxford university press.

\bibitem[{Wilks and Stacey(2016)}]{wilks2016improvements}
Wilks, T.; and Stacey, W. 2016.
\newblock Improvements to an ion orbit loss calculation in the tokamak edge.
\newblock \emph{Physics of Plasmas}, 23(12).

\end{thebibliography}
%%%%%%%%%%%%%%%%%%%%%%%%%%%%%%%%%%%%%%%%
\end{document}